\documentclass[pra,twocolumn,superscriptaddress,eqsecnum,showpacs,floatfix]{revtex4}
\usepackage[normalem]{ulem}
\usepackage{amsmath,amssymb,amsfonts,times,graphicx}
\usepackage[ps2pdf,bookmarks=true,colorlinks,linkcolor=red,urlcolor=blue,citecolor=blue]{hyperref}
\usepackage[usenames]{color}
\usepackage{dcolumn} 
\usepackage{bm}

\def \k{{\mathbf{k}}}
\def \q{{\mathbf{q}}}
\def \p{{\mathbf{p}}}

\begin{document}
\title{Polaron to molecule transition in a strongly imbalanced Fermi gas}

\author{M. Punk}
\affiliation{Physik Department, Technische Universit\"at M\"unchen, James-Franck-Strasse, D-85748 Garching, Germany}

\author{P. T. Dumitrescu}
\affiliation{Jesus College, University of Cambridge, Cambridge CB5 8BL, UK}

\author{W. Zwerger}
\affiliation{Physik Department, Technische Universit\"at M\"unchen, James-Franck-Strasse, D-85748 Garching, Germany}

\date{\today}

\begin{abstract}
A single down spin Fermion with an attractive, zero range interaction 
with a Fermi sea of up-spin Fermions forms a polaronic quasiparticle.
The associated quasiparticle weight vanishes beyond a 
critical strength of the attractive interaction, where a many-body bound state
is formed. From a variational wavefunction in the molecular limit, we
determine the critical value for the polaron to molecule transition.
The value agrees well with the diagrammatic Monte Carlo results of
Prokof'ev and Svistunov and is consistent with recent rf-spectroscopy
measurements of the quasiparticle weight by Schirotzek 
et. al. In addition, we calculate the contact coefficient of the 
strongly imbalanced gas, using the adiabatic theorem of Tan 
and discuss the implications of the polaron to molecule transition
for the phase diagram of the attractive Fermi gas at finite imbalance.

\end{abstract}


\maketitle

\section{Introduction}

The physics of single particles immersed in an environment is ubiquitous in physics.
It appears, for example, in the large polaron problem where a single electron is dressed by 
its interaction with phonons \cite{Feynman72} or in models for dissipation
and decoherence in quantum mechanics  \cite{Caldeira81,Leggett87}.
In recent years, new directions for exploring quantum many-body problems have 
been opened through ultracold atoms \cite{Bloch08}. In particular, for degenerate Fermi gases,
the interaction strength can be tuned over a wide range using Feshbach resonances.
This allows to study impurity problems in a fermionic environment. A specific example 
is a gas of fermionic $^6$Li, where the  two lowest hyperfine states are populated in a 
highly imbalanced situation. For this system, recent experiments have shown that the 
minority atoms ('down spins') apparently form a liquid of  quasiparticles \cite{Schirotzek09}. 
Due to the strong attractive interaction to the up-spin Fermi sea, the associated quasiparticle 
weight, as  determined from a sharp peak in the rf-spectrum, is found to vanish 
beyond a critical interaction strength. This transition may be interpreted 
as one, in which a single $\downarrow$-Fermion immersed in sea of $\uparrow$-Fermions
can no longer propagate as a quasiparticle but forms a 
many-body bound state with the Fermi sea.
The existence of such a transition has been 
predicted by Prokof'ev and Svistunov \cite{Prokofev1,Prokofev2}.
Using a novel diagrammatic Monte Carlo method, they have shown that
for strong attractive interactions, a molecular state is energetically 
favored compared to one in which the single down-spin forms
a polaronic quasiparticle in the up-spin Fermi sea. 
In the present work, we analyze this problem by a simple variational
wavefunction. It provides an analytically tractable model for 
the physics on the molecular side, thus complementing
the variational description put forward by Chevy \cite{Chevy06} 
for the polaronic quasiparticle. Our wavefunction gives a ground state energy 
that matches perfectly the results of the diagrammatic Monte Carlo method.
Moreover, it describes correctly the three-body physics of repulsive 
atom-dimer interactions in the deep molecular limit and
has zero residue for the down-spin Green function. The variational wave 
function is also used to determine the saturation field $h_s$  beyond which 
a two component Fermi gas is fully polarized and the behavior of the so-called
contact coefficient introduced by Tan \cite{Tan1} in the limit of strong imbalance.

\section{The Fermi Polaron and its Quasiparticle Weight}
\label{section_2}

A simple variational wavefunction for the (N+1)-particle problem of a single 
down-spin Fermion immersed in a sea of spin-up Fermions has been 
introduced by Chevy \cite{Chevy06}. It is based on an expansion 
up to single particle-hole excitations around the unperturbed Fermi sea 
\begin{equation}
| \psi_0 \rangle = \Big( \phi_0 \, d^\dagger_{\mathbf{0}} + \sideset{}{'}\sum_{\k, \q}
 \phi_{\k \q} \,  d^\dagger_{\mathbf{q-k}} u^\dagger_{\mathbf{k}} u_{\mathbf{q}} \Big)
 | FS^N_\uparrow \rangle \ .
\label{chevy}
\end{equation}
Here and in the following sums on $\k$ and $\q$ with a prime are restricted to
 $k>k_F$ and $q<k_F$, respectively. Moreover, $|FS^N_\uparrow\rangle$ is the N-particle
 Fermi sea and the creation operators of up- and down-Fermions with momentum $\k$
 are denoted by $u^\dagger_\k$ and $d^\dagger_\k$. Despite the restriction
 to single particle-hole excitations, which is difficult to justify for the relevant case
 of zero range interactions that can create particle-hole pairs at arbitrary momentum,
 Monte Carlo calculations show that the ansatz \eqref{chevy} 
 gives a ground state energy that is very accurate, in particular at unitarity, where
 the scattering length $a$ is infinite \cite{Prokofev1,Prokofev2}. The reason
 why the leading term in an expansion in the number of particle-hole
 excitations gives very good results for the ground state energy can be traced 
 back to the decoupling of higher order terms for vanishing 
 hole momenta $\mathbf{q}=0$ \cite{Combescot1},
i.e.\ contributions with more than one particle hole excitation interfere destructively.

The wavefunction \eqref{chevy} describes the added down-spin as a quasiparticle
dressed by its interaction with the up-spin Fermi sea. The virtual cloud of particle-hole 
excitations leads to a quasiparticle energy
\begin{equation}
E(\mathbf{p})=A\varepsilon_F +\frac{\mathbf{p}^2}{2m^{\star}} +\ldots
\label{E(p)}
\end{equation}
at low momenta $|\mathbf{p}|\ll k_F$ that contains a 'binding energy' $A\varepsilon_F<0$ 
of a single down-spin to the Fermi sea and an effective mass $m^{\star}$ \cite{Lobo06}.
Here, the Fermi energy is defined by $\varepsilon_F=k_F^2/(2 m)$ (we use $\hbar=1$ throughout the paper) with a Fermi momentum $k_F$ that is related to the up-spin density by  
the standard relation $n_\uparrow=k_F^3/(6 \pi^2)$ for a single component 
Fermi gas. Since we are interested in the limit of vanishing down-spin density 
$n_{\downarrow}\to 0$, these are the relevant energy and momentum scales. 
The dimensionless coefficient $A$ and the effective mass $m^{\star}$ have been 
determined from variational Monte Carlo calculations at the unitarity point \cite{Lobo06}
and from a T-matrix approximation at arbitrary values of the 
dimensionless interaction strength $v=1/(k_Fa)$ \cite{Combescot2}.  
Very recently they have also been measured experimentally, giving 
$A\approx -0.64(7)$  \cite{Schirotzek09} and $m^{\star}/m=1.17(10)$ 
\cite{Salomon09} at unitarity, in rather good agreement with the theoretical
predictions.  

From a many-body point of view, the criterion that a single added down-spin 
is indeed a proper quasiparticle can be expressed by defining the 
quasiparticle residue $Z_\downarrow$ from the long-time limit 
\begin{equation}
Z_{\downarrow}=\lim_{t\to\infty}\vert G_{\downarrow}(\mathbf{p}=0,t)\vert\ne 0
\label{G(t)}
\end{equation}
of the down-spin Green-function at zero momentum. Within the variational 
wavefunction \eqref{chevy} this residue is simply given by the probability 
$Z_\downarrow = |\phi_0|^2$ that an added down-spin at momentum 
$\mathbf{p}=0$ is not mixed with plane waves at nonzero momenta
$\mathbf{q}-\mathbf{k}\ne 0$ through particle-hole excitations. 
The fact that the coefficient $|\phi_0|^2$ of the Chevy wavefunction coincides with 
the quasiparticle weight can be derived formally by noting that the ansatz 
\eqref{chevy} is equivalent to a non-selfconsistent T-matrix approach for 
the down-spin Green function,  which  sums
the particle-particle ladder for the vertex part $\Gamma(\k,\omega)$ \cite{Combescot2}.
It is then straightforward to see that $|\phi_0|^2=\vert 1-\partial_{\omega}\Sigma|_{\omega=0}\vert^{-1}$ 
coincides with the standard definition of the quasiparticle weight via the 
energy derivative of the down-spin self energy $\Sigma(\mathbf{p},\omega)$
at zero frequency and momentum. The numerical value of $Z_\downarrow$ 
at unitarity $v=0$ is $Z_\downarrow(v\!=\!0)\simeq 0.78$ within the Chevy
ansatz. This is much larger than the experimentally observed value $Z_\downarrow=0.39(9)$ 
which is likely to be a lower bound, however \cite{Schirotzek09}.  Smaller values
$Z_\downarrow(v=0)\!=\!0.47$ of the quasiparticle weight at unitarity are 
found from a $1/N$-expansion of the attractive fermion problem at strong
imbalance, which is equivalent to a non-selfconsistent T-matrix approximation
with the bare chemical potential \cite{Sachdev08}.

\begin{figure}
\includegraphics[width=0.95 \columnwidth]{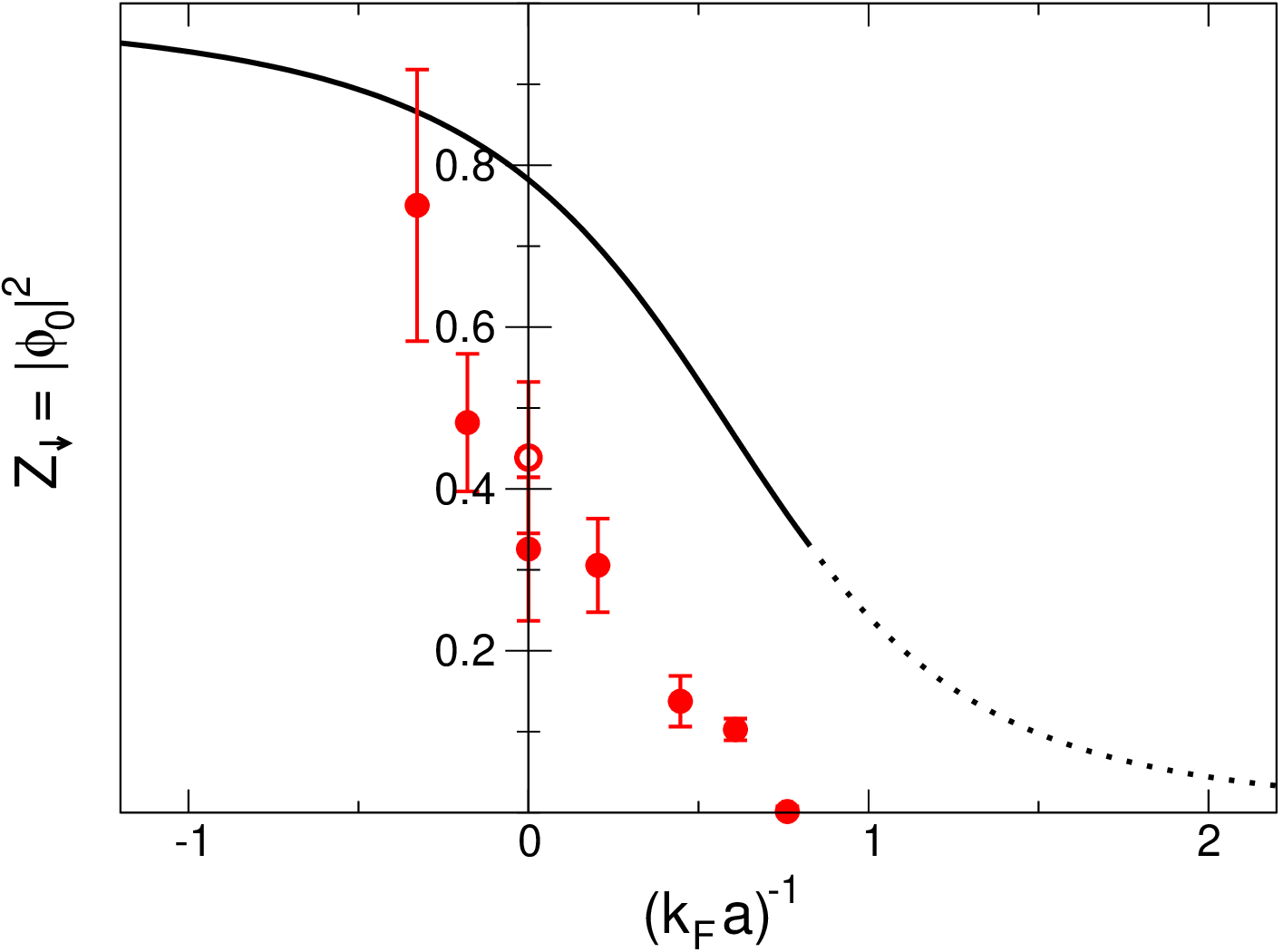}
\caption{(Color online) Quasiparticle residue $Z_\downarrow$ of the minority Fermion as
 function of $(k_F a)^{-1}$, calculated using Chevy's variational ansatz \eqref{chevy}.
 In the regime where the ansatz \eqref{chevy} breaks down, $Z_\downarrow$ is drawn as dotted line.
 The red dots correspond to the experimentally measured quasiparticle residue from the MIT group
 \cite{Schirotzek09} at a minority concentration of 5\%. }
\label{fig:Zchevy}
\end{figure}

In Fig. \ref{fig:Zchevy} we show the quasiparticle residue $Z_\downarrow$ for
 the minority Fermion as a function of $v=1/(k_F a)$ within the ansatz
 \eqref{chevy} in comparison with the recent experimental results 
 \cite{Schirotzek09}. Apparently, the expansion up to single particle-hole
 excitations considerably overestimates the quasiparticle residue even though it 
 gives reliable results for the ground state energy. A much more basic 
 shortcoming of the ansatz \eqref{chevy}, however, appears if one
 considers the BEC-limit $v\gg 1$. Indeed, the ansatz predicts a finite
 value of  $Z_\downarrow$ at arbitrary interaction strengths, even in the 
 deep molecular limit. In this limit, however, an added down-spin will 
 form a bound state with one of the up-spin fermions and can no longer
 propagate as a coherent quasiparticle. One thus expects that  $Z_\downarrow$
 vanishes identically beyond a critical strength $v_M>0$ of the interaction,
 consistent with the experimental findings \cite{Schirotzek09}.  
 It is important to note, that the formation of
 the bound state is a genuine many-body effect at any finite density of
 the up-spin Fermi sea. Indeed, the binding is of a two-body nature only in the trivial limit 
 $v\gg 1$ (that is, effectively, for $k_F\to 0$), where the bound state is
 formed with a single up-spin Fermion. By contrast, just beyond the critical value 
 $v_M$, the down-spin is effectively compensated by forming a singlet with
 many up-spin Fermions, somewhat similar to the physics of a localized
 Kondo-spin interacting antiferromagnetically with a sea of conduction electrons
 at temperatures much below the Kondo-temperature \cite{Wilson75}.  
 Note, however, that in the Kondo problem the impurity spin is not fixed and 
 the transition from an uncompensated spin to an effective singlet state 
 appears as a continuous crossover from high to low temperatures. 
 In the present problem, instead, there is a discontinuous 
 transition in the ground state as a function of the attractive coupling $v$.
 
 An indication that the variational wavefunction \eqref{chevy} is not applicable
 for strong attractive interactions is provided by considering the 
 Thouless criterion for a superfluid instability in which up-and down-spins
are paired in an s-wave superfluid \cite{Thouless61}. Evaluating the
relevant vertex function within the T-matrix approximation for arbitrary values of
the down-spin chemical potential $\mu_{\downarrow}$, it is found that the 
Thouless criterion $\Gamma^{-1}(\k=0,\omega=0)=0$ leads to a critical value
for $\mu_{\downarrow}$ that is below the value $\mu_{\downarrow}=E(\mathbf{p}=0)$
obtained from the ground state energy of the variational state \eqref{chevy} 
provided that $v\geq 1.27$. A second argument that indicates the breakdown
of the ansatz  \eqref{chevy} in the regime $v\gg 1$ is the behavior of the ground state 
energy. Indeed, in a systematic expansion in powers of the scattering length $a\to 0^+$ 
the ground state energy of a single added down-spin relative to the free
up-spin Fermi sea is expected to be of the form 
\begin{equation}
E \overset{a \rightarrow 0^+}{=} E_b-\varepsilon_F + g_{ad} n_\uparrow + 
\mathcal{O}(a^2)\, .
\label{asympen}
\end{equation}
Its leading contribution is just the molecular binding energy $E_b=-1/(m a^2)<0$.
The contribution $\varepsilon_F$ of order $a^0$ accounts for the removal of one 
$\uparrow$-Fermion from the Fermi-sea that is required for the formation of the 
molecule. The last term, of order $a$ is the mean field repulsion between the molecule and the 
Fermi-sea. Its interaction strength $g_{ad}= 3 \pi a_{ad}/m$ is related to the 
exact atom-dimer scattering length $a_{ad}=1.18\,  a$ that has first been 
calculated in connection with neutron-deuteron scattering \cite{Skorniakov57} 
(for a recent derivation in the cold gas context see Petrov et. al. \cite{Petrov05}). 
It turns out, however, that the variational ansatz Eq. \eqref{chevy} leads to
 $E \, \overset{a\rightarrow 0^+}{=} \, E_b -\varepsilon_F/2 + \mathcal{O}(a)$ which is 
 too high by $\varepsilon_F/2$ compared to the exact asymptotics \eqref{asympen}. 
 The reason for this discrepancy
can be seen easily from the structure of Chevy's wavefunction. On the BEC side, the
 dominant contribution comes from the $q=0$ terms, i.e.
\begin{equation}
\sideset{}{'}\sum_\mathbf{k} \phi_{\mathbf{k, 0}} \,  d^\dagger_{\mathbf{-k}}
 u^\dagger_{\mathbf{k}} u_{\mathbf{0}}  | FS^N_\uparrow \rangle
\end{equation}
describing the molecule formation of an up- and a down-spin with opposite momenta.
 This contribution is not optimal however, since it creates a hole in the center of
 the $\uparrow$-Fermi-sphere. Energetically, it would be favorable to replace
 $u_{\mathbf{0}}  | FS^N_\uparrow \rangle$ with a (N-1)-particle Fermi sea
 $| FS^{N-1}_\uparrow \rangle$, leading to a ground state energy that is lower by
 $\varepsilon_F$. Within the ansatz \eqref{chevy}, this would require terms with an
 arbitrary number of particle-hole excitations in order to reshuffle the Fermi-sea
 in such a way that the hole vanishes.

\section{Variational ansatz in the molecular regime}

In order to describe the physics of bound state formation in the regime $v\gg 1$, we propose a 
variational ansatz for the (N+1)-body problem that complements
the ansatz \eqref{chevy} describing a Fermi polaron with a finite quasiparticle residue. 
Our ansatz gives the exact behavior \eqref{asympen} of the ground state energy in the BEC-limit
up to linear order in $a$.  The associated variational wavefunction 
\begin{eqnarray}
| \psi_0 \rangle &=& \Big( \sideset{}{'}\sum_{\mathbf{k}} \xi_\mathbf{k} \,
 d^\dagger_{\mathbf{-k}} u^\dagger_{\mathbf{k}} \notag \\  
&& + \sideset{}{'}\sum_{\mathbf{k', k, q}} \xi_{\mathbf{k' k q}} \, 
d^\dagger_{\mathbf{q-k-k'}} u^\dagger_{\mathbf{k'}} u^\dagger_{\mathbf{k}} 
u_{\mathbf{q}}  \Big) | FS^{N-1}_\uparrow \rangle
\label{Varansatz2}
\end{eqnarray}
is a natural generalization of the Chevy ansatz and is constructed by adding a 
 $(\uparrow, \downarrow)$-pair to a $(N-1)$-particle
 Fermi sea of $\uparrow$-Fermions, together with the leading term in an expansion
 in particle-hole excitations.
Again, sums on $\k, \k'$ and $\q$ are restricted to $k, k' >k_F$ and $\q<k_F$,
 respectively. The first term accounts for the formation of the molecule in the
 presence of the $\uparrow$-Fermi sea and gives the correct next-to-leading-order
 ground state energy in the BEC-limit, avoiding the problem of creating a hole in the
 $\uparrow$-Fermi sea. The single particle-hole excitation in the second term describes
 the leading order contribution to the interaction of the dimer with the Fermi sea
 apart from Pauli-blocking effects, that are already accounted for in the first term.
 An important feature brought about by the inclusion of the second term in 
 Eq. \eqref{Varansatz2} is that it amounts to an
 exact treatment of the three-particle problem.  Indeed, as is shown in detail
 in the Appendix, the set of coupled equations (\ref{e1})-(\ref{e4}) that determine 
 the coefficients of the variational many-body wavefunction reduce, in the three-particle limit,
 precisely to the integral equation for the exact solution of the three-body problem
 by Skorniakov and Ter-Martirosian \cite{Skorniakov57}. As a result,  the exact 
 atom-dimer scattering length $a_{ad}=1.18a$ appears in the asymptotic behavior of the ground state
 energy \eqref{asympen}, giving rise to the correct next-to-next-to-leading order
 behavior of the ground state energy in the BEC-limit.

Obviously, the ansatz \eqref{Varansatz2} is not capable of describing the whole range
 of scattering lengths correctly. In particular, it does not capture the weak coupling limit
 $a\rightarrow 0^-$. Indeed, the $\downarrow$-Fermion in the first term is
 always added at momenta $k>k_F$, leading to a ground state energy that is too high
 by $\varepsilon_F$ in the weak coupling limit.  Our ansatz \eqref{Varansatz2} is
 therefore complementary to the Chevy wavefunction \eqref{chevy}, which 
correctly describes the situation at weak coupling up to and slightly beyond the
unitarity limit.

From a physical point of view, the two variational wavefunctions \eqref{chevy} and
 \eqref{Varansatz2} characterize very different ground states.
 Chevy's ansatz describes a Fermi polaron with a finite quasiparticle residue, 
 which allows to build a normal Fermi liquid at a finite concentration of the 
 down-spin Fermions, provided that interactions between the quasiparticles 
 have no attractive channels (see section IV. below). 
By contrast, the wavefunction \eqref{Varansatz2} describes a bosonic
molecule interacting with a Fermi sea. At a finite concentration $n_{\downarrow}\ne 0$,
the resulting ground state is expected to be a superfluid, coexisting with unpaired
up-spin Fermions. The critical coupling $v_M$,  where the ground state energies 
of the two variational wavefunctions intersect is thus expected to separate 
a normal fluid from a superfluid ground state of the attractive Fermi gas 
in the limit of very strong imbalance. 

The variational ansatz \eqref{Varansatz2} is based on a single channel model
that describes the attractive interactions between both spin states. For computational purposes, 
however,  it turns out to be easier to start from the more general two-channel model, which is
 defined by the Hamiltonian
\begin{eqnarray}
H &=& \sum_\p \left(\frac{\varepsilon_\p}{2} +\nu_0 \right) b_\p^\dagger b_\p
 + \sum_{\p, \sigma} \varepsilon_\p c^\dagger _{\p,\sigma} c _{\p,\sigma} \notag \\
& & + \frac{g_0}{\sqrt{V}} \sum_{\p',\p} \left( b^\dagger_\p c_{\p-\p',\uparrow}
 c_{\p',\downarrow} \ + \ h.c.\right) \ .
\label{H2channel}
\end{eqnarray}
Here, $b^\dagger_\p$ denotes the bosonic creation operator of a molecule with momentum
 $\p$ and $c^\dagger _{\p,\sigma}$ are the fermionic creation operators for the two
 species $\sigma=\uparrow, \downarrow$. The free particle dispersion is denoted by
 $\varepsilon_\p=p^2/(2m)$ and the factor $1/2$ in the first term accounts for the factor two
in the molecule to single Fermion mass ratio. The bare values of the detuning $\nu_0$ and 
the Feshbach coupling strength $g_0$ can be related to the physical s-wave 
scattering length $a$ and the interaction range $r_0$ via \cite{Bruun04}
\begin{eqnarray}
\frac{\nu_0}{g_0^2} &=& -\frac{m}{4 \pi a} +\frac{1}{V} \sum_\p 
\frac{1}{2 \varepsilon_\p} \ , \\
r_0 &=& - \frac{8 \pi}{g_0^2 m^2} \ .
\end{eqnarray}
The two-channel Hamiltonian \eqref{H2channel} is equivalent to a single channel model
in the interesting limit of zero-range interactions $r_0 \rightarrow 0$
 (i.e. for broad Feshbach resonances), as can be seen easily by integrating out the
 bosonic degrees of freedom.

The corresponding variational ansatz to \eqref{Varansatz2} in the two-channel model has
 two additional terms ($\sim \eta_0, \, \eta_{\k \q}$) where the closed-channel state is
 occupied
\begin{eqnarray}
| \psi_0 \rangle &=& \Big( \eta_0 \, b^\dagger_\mathbf{0} + \sideset{}{'}\sum_{\mathbf{k}}
 \xi_\mathbf{k} \, d^\dagger_{\mathbf{-k}} u^\dagger_{\mathbf{k}} +
 \sideset{}{'}\sum_{\k,\q} \eta_{\k \q} \, b^\dagger_{\q-\k} u^\dagger_\k u_\q \notag \\
& &+ \sideset{}{'}\sum_{\mathbf{k', k, q}} \xi_{\mathbf{k' k q}} \, 
d^\dagger_{\mathbf{q-k-k'}} u^\dagger_{\mathbf{k'}} u^\dagger_{\mathbf{k}} u_{\mathbf{q}}
  \Big) | FS^{N-1}_\uparrow \rangle \ .
\label{ansatz2}
\end{eqnarray}
Calculating the expectation value $\langle \psi_0 | \hat{H}-E | \psi_0 \rangle$, taking the
 derivatives with respect to the infinite set of variational parameters
  $\eta_0, \, \xi_\mathbf{k}, \, \eta_{\k \q}, \, \xi_{\mathbf{k' k q}}$ and setting them
 equal to zero leads to the following set of coupled equations
\begin{widetext}
\begin{eqnarray}
\left(E+\varepsilon_F-\nu_0 \right) \eta_0 &=& - \frac{g_0}{\sqrt{V}} 
\sideset{}{'}\sum_\k \xi_\k \label{e1}\\
\left(E+\varepsilon_F-2 \varepsilon_\k \right) \xi_\k &=& - \frac{g_0}{\sqrt{V}}
 \eta_0 + \frac{g_0}{\sqrt{V}} \sideset{}{'}\sum_\q \eta_{\k \q} \label{e2} \\
\left(E+\varepsilon_F-\nu_0-\frac{\varepsilon_{\q-\k}}{2}-\varepsilon_\k+\varepsilon_\q
 \right) \eta_{\k \q} &=& \frac{g_0}{\sqrt{V}} \xi_\k - \frac{2 g_0}{\sqrt{V}} 
\sideset{}{'}\sum_{\k'} \xi_{\k' \k \q} \label{e3}\\
\left(E+\varepsilon_F-\varepsilon_{\q-\k-\k'}-\varepsilon_{\k'}-\varepsilon_\k+
\varepsilon_\q \right) \xi_{\k' \k \q} &=& -\frac{g_0}{2 \sqrt{V}} \left( \eta_{\k \q}
 - \eta_{\k' \q} \right) \label{e4}
\end{eqnarray}
\end{widetext}
Note that the ground state energy $E$ is measured with respect to the N-particle Fermi sea,
 which explains the occurrence of the $\varepsilon_F$ terms in the above equations.
Moreover, using the N-particle Fermi sea as reference scale, the ground state energy $E$ is equivalent to the chemical potential $\mu_\downarrow \equiv E$ of the single down-spin.

\subsection{No particle-hole excitation}

Neglecting for a moment the contribution of particle-hole excitations in \eqref{ansatz2},
 i.e. setting $\eta_{\k \q} = \xi_{\k' \k \q}=0$, the ground-state energy 
 is determined by the equations \eqref{e1} and \eqref{e2} alone.
Performing the integrations and taking the zero-range limit $r_0 \rightarrow 0$, they 
reduce to a simple transcendental equation
\begin{equation}
\frac{\pi}{2 k_F a}=1+\sqrt{-\frac{E+\varepsilon_F}{2 \varepsilon_F}} \arctan \left( 
\sqrt{-\frac{E+\varepsilon_F}{2 \varepsilon_F}} \right)\, .
\label{thouless}
\end{equation}
In the BEC-limit $a \rightarrow 0^+$, Eq. \eqref{thouless} gives rise to a ground state
 energy of the form \eqref{asympen}. The associated atom-dimer scattering length, 
however,  is given by its value $a_{ad}^{Born}=(8/3) \, a$ in the Born approximation. 
More generally, it turns out that Eq. \eqref{thouless} is exactly equivalent to the Thouless-criterion
 $\Gamma^{-1}(\mathbf{k}=0,\omega=0)=0$ if the vertex function is calculated
 within a non-selfconsistent T-matrix approach where only the particle-particle ladder is 
 summed, as discussed in section \ref{section_2}. The resulting ground state energy 
 is below that of the Fermi polaron if $v\geq 1.27$.   

\subsection{Full variational treatment}

In the general case $\eta_{\k \q} \neq0, \ \xi_{\k' \k \q} \neq 0$, the equations
 (\ref{e1})-(\ref{e4}) can be reduced to a single homogeneous Fredholm equation of the
 second kind for the variational parameters $\eta_{\k \q}$ in the thermodynamic limit
 (again, the zero range limit has been taken already)
\begin{equation}
\frac{1}{V^2} \sideset{}{'}\sum_{\k', \q'} K(E; \k,\q ; \k', \q') \eta_{\k' \q'} = 0 \, .
\label{intequ}
\end{equation}
The associated Kernel $K(E; \k,\q ; \k', \q')$ is given by
\begin{equation}
K = \frac{V \delta_{\k,\k'}}{E_\k}-\frac{1}{\gamma E_\k E_{\k'}}-
\frac{V \delta_{\q,\q'}}{E_{\k'\k\q}}-\alpha_{\k \q} V^2  \delta_{\k,\k'}
 \, \delta_{\q,\q'}
\end{equation}
with
\begin{eqnarray}
E_\k &\doteq& E+\varepsilon_F-2 \varepsilon_\k \\
E_{\k' \k \q} &\doteq& E+\varepsilon_F-\varepsilon_{\q-\k-\k'}-\varepsilon_\k-
\varepsilon_{\k'}+\varepsilon_\q \label{coeff_E3}\\
\alpha_{\k \q} &\doteq& -\frac{\nu_0}{g_0^2}-\frac{1}{V} \sideset{}{'}\sum_{\k'}
 \frac{1}{E_{\k' \k \q}} \label{coeff_alpha}\\
\gamma &\doteq& \frac{\nu_0}{g_0^2}+\frac{1}{V} \sideset{}{'}\sum_\k\frac{1}{E_\k}
\end{eqnarray}
Due to the isotropy of the system, the variational parameters
 $\eta_{\k \q} \equiv \eta(k,q,\cos \theta_{\k \q})$ depend only on the magnitudes 
 of the two momenta $\k$ and $\q$ and the
 angle between them. This allows Eq. \eqref{intequ} to be reduced to a
 three dimensional integral equation. 

The ground state energy $E$ is now simply obtained by the condition that the Fredholm
 determinant of the kernel $K$ vanishes. We evaluate the Fredholm determinant numerically
 by discretizing the integral equation using a Gauss-Legendre quadrature and calculating
 the determinant of the corresponding linear equation system. The order of the quadrature
 for the $k$, $q$ and $\cos \theta_{\k \q}$ integral were chosen as 11, 11 and 4, leading
 to an error of $\sim 10^{-4}$ of the ground state energy at the unitarity point
 $a\rightarrow \infty$, where the convergence is slowest.

\begin{figure}
\includegraphics[width=\columnwidth]{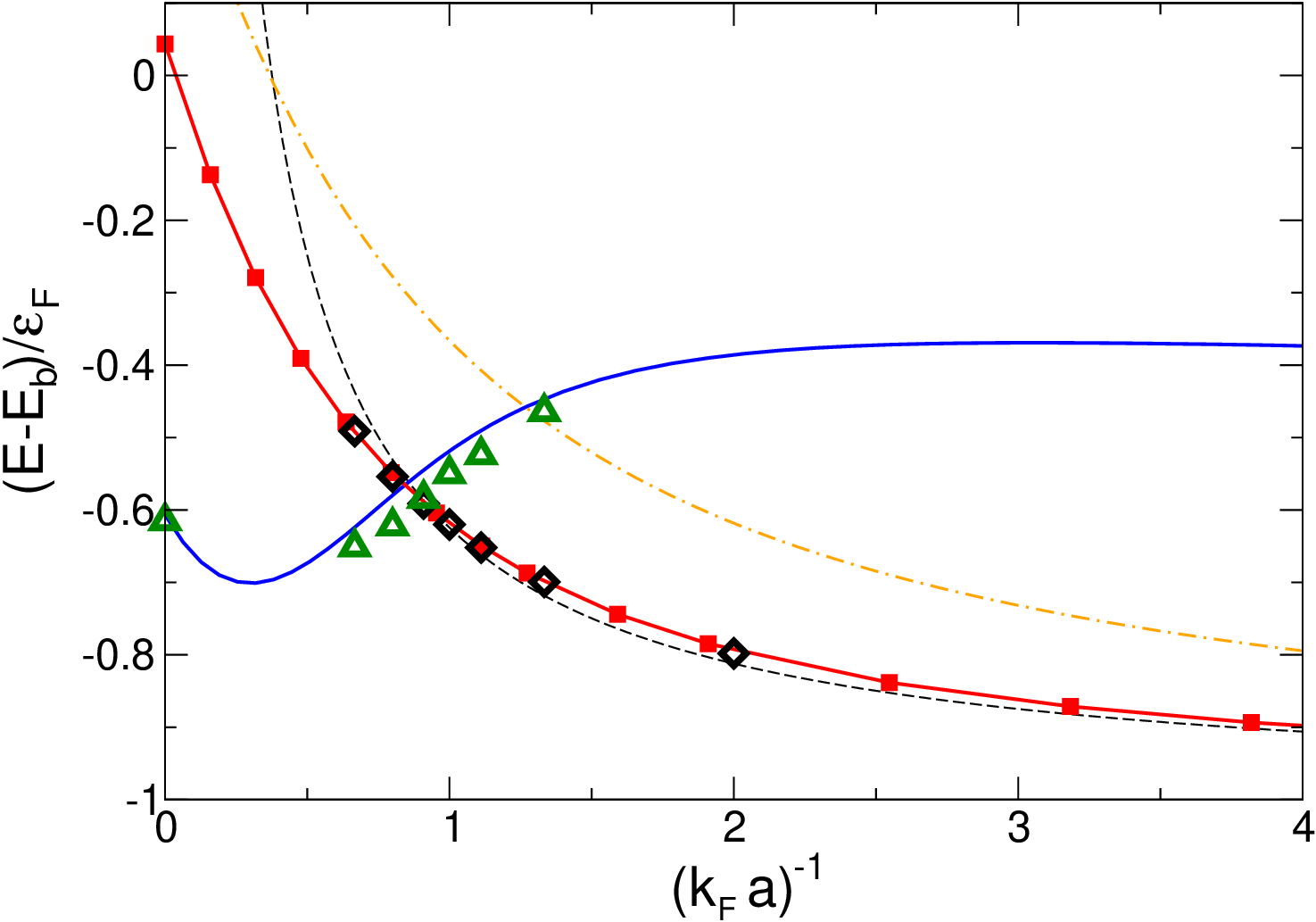}
\caption{(Color online) Ground state energy $E-E_b$ (binding energy $E_b=-1/(ma^2)$ subtracted)
 in units of the Fermi energy $\varepsilon_F$ as function of $(k_F a)^{-1}$. Blue solid line:
 Chevy's ansatz \eqref{chevy}; red line with full squares: ansatz Eq.
 \eqref{Varansatz2}; black dashed line: BEC-asymptotics \eqref{asympen}; orange dash-dotted
 line: Thouless criterion \eqref{thouless}. The open black diamonds and green triangles correspond
 to the QMC results for the molecule- and the polaron energy from Prokof'ev and Svistunov \cite{Prokofev1}.
}
\label{fig:GSen}
\end{figure}

The ground state energy as function of $(k_F a)^{-1}$ is shown in Fig. \ref{fig:GSen}.
 Apparently, our ansatz \eqref{Varansatz2} leads to a ground state energy that is 
 below that of Chevy's ansatz
 for interaction strengths larger than $(k_F a)_\text{M}^{-1} = 0.84$.
 This value is in good agreement with the diagrammatic Monte-Carlo results by 
 Prokof'ev and Svistunov \cite{Prokofev1, Prokofev2},
 who obtain $(k_F a)_\text{M}^{-1}=0.90$. In fact, the small discrepancy is entirely 
 due to the fact that we intersect our molecular ground state energy with that
 obtained using the Chevy wavefunction, which is not precise near $v_M$.  
The Monte Carlo results in turn give better values for
 the polaron energy, shifting the intersection slightly towards the BEC regime,
as can be seen in Fig. \ref{fig:GSen}.. Yet, as far as the molecular ground state energy 
is concerned, our results agree perfectly with the Monte Carlo data, down to the 
smallest coupling $v\simeq 0.6$ where they have been calculated. 

It is interesting to note, that the approximation $\q=0$ in the wavefunction
 \eqref{ansatz2} (i.e. pinning the hole-wavevector at zero momentum), leads to a ground
 state energy that differs from the calculation with the full wavefunction by at most  3\%
 in the regime $(k_F a)^{-1}>0.84$, where the ansatz is valid. The situation thus 
 appears similar to that in the polaron case, where Combescot et al. \cite{Combescot1}
 have shown that an expansion in hole-wavevectors works very well for Chevy's 
 ansatz at unitarity.

\subsection{Quasiparticle Residue}

We now show that the quasiparticle residue $Z_\downarrow$ of the $\downarrow$-Fermion,
 which can be thought of as a kind of order parameter of the transition from the polaron to the
 molecular state, vanishes identically in the thermodynamic limit for the variational
 wavefunction \eqref{ansatz2}, that gives a lower ground state energy on the molecular side of 
 the critical coupling $v_M$.  
 
 Since the variational ground state wavefunction does not allow to calculate the
 full down-spin Green Function, the definition \eqref{G(t)} of the quasiparticle residue
 is not applicable. Instead, we use the standard connection between $Z_\downarrow$
 and the jump in the momentum distribution at the Fermi momentum $k_{F\downarrow}$,
 and the latter is zero in the limit of a single down-spin.  
The momentum distribution of the $\downarrow$-Fermion within the variational ansatz
 \eqref{ansatz2} is given by
\begin{equation}
n^{\downarrow}_{\p}= \left| \xi_{\p}\right|^2+ 2 \sideset{}{'}\sum_{\k' \k \q} 
\left| \xi_{\k' \k \q}\right|^2 \delta_{\p,\q-\k'-\k}
\end{equation}
and is normalized via
\begin{equation}
1=\sum_\p n^{\downarrow}_{\p} = \sideset{}{'}\sum_\k \left| \xi_{\k}\right|^2 
+ 2 \sideset{}{'}\sum_{\k' \k \q} \left| \xi_{\k' \k \q}\right|^2 \ .
\end{equation}
The normalization condition requires the coefficients to scale with the system 
volume as $\xi_\k \sim 1/\sqrt{V}$ and $\xi_{\k' \k \q} \sim 1/V^{3/2}$.
Since an upper bound to the quasiparticle residue $Z_\downarrow$ is given 
by the momentum distribution at $\p=0$ and $\xi_\p \equiv 0$ for $p<k_F$, we find that
\begin{equation}
Z_\downarrow \leq n^{\downarrow}_{\p=0} = 2 \sideset{}{'} \sum_{\k' \k \q} \left| 
\xi_{\k' \k \q}\right|^2 \delta_{\q,\k'+\k} \sim \frac{1}{V} \, .
\end{equation}
As a result, the quasiparticle residue $Z_\downarrow$ of the molecular wavefunction 
scales inversely with the volume of the system and thus vanishes in the thermodynamic
 limit. This is in contrast to Chevy's wavefunction, where
 $Z_\downarrow = |\phi_0|^2$ is always finite. The two wavefunctions \eqref{chevy} and \eqref{Varansatz2} therefore indeed describe qualitatively different ground states.
 In particular, no sharp peak is expected in the minority rf-spectrum
 at coupling strengths $v>v_M$, consistent with the experimental observation
 \cite{Schirotzek09}.

In the $\q=0$ approximation, which captures the essential properties of the variational
 ansatz \eqref{Varansatz2}, the quasiparticle residue $Z_\downarrow$ in fact  
vanishes identically. Indeed,  
\begin{equation}
Z_\downarrow \leq 2 \sideset{}{'} \sum_{\k} \left| \xi_{-\k \k 0}\right|^2 
= 0  \ ,
\end{equation}
since, as can be seen from Eq. \eqref{e4}, the coefficients
 $\xi_{-\k \k 0} \propto \eta_{\k0}-\eta_{-\k0} = 0$ vanish because $\eta_{\k0}$ only
 depends on the length of $\k$.

\section{Contact coefficient and Phase Diagram}

The analysis of the polaron to molecule transition in the previous
section leaves two important questions open: what is the nature of
the transition and what are its implications for the phase diagram of the 
strongly imbalanced gas? Now for the case of a single down-spin in
an up-spin Fermi sea, the transition from a polaronic to a molecular state 
is a first order transition, where the quasiparticle residue $Z_{\downarrow}$ 
exhibits a discontinuous jump from a finite value to zero at the critical 
coupling $v_M\simeq 0.9$. This is a result of the fact that the energies
of the two ground states, which have different quantum numbers,
cross with a finite slope at $v_M$ (see Fig. \ref{fig:GSen}.). It is 
important to note that this crossing is not an artefact of extending 
the different variational states beyond their domain of validity.
Indeed, as shown by Prokof'ev and Svistunov \cite{Prokofev1, Prokofev2},
both the polaronic and the molecular state exist as stable excitations 
for $v>v_M$ or $v<v_M$ respectively because the phase space for
decay vanishes linearly with the magnitude of the energy difference. 
Both states are thus reachable as metastable configurations coming
from the weak coupling or the molecular side, as expected for a first order
transition.

\begin{figure}
\includegraphics[width=0.95 \columnwidth]{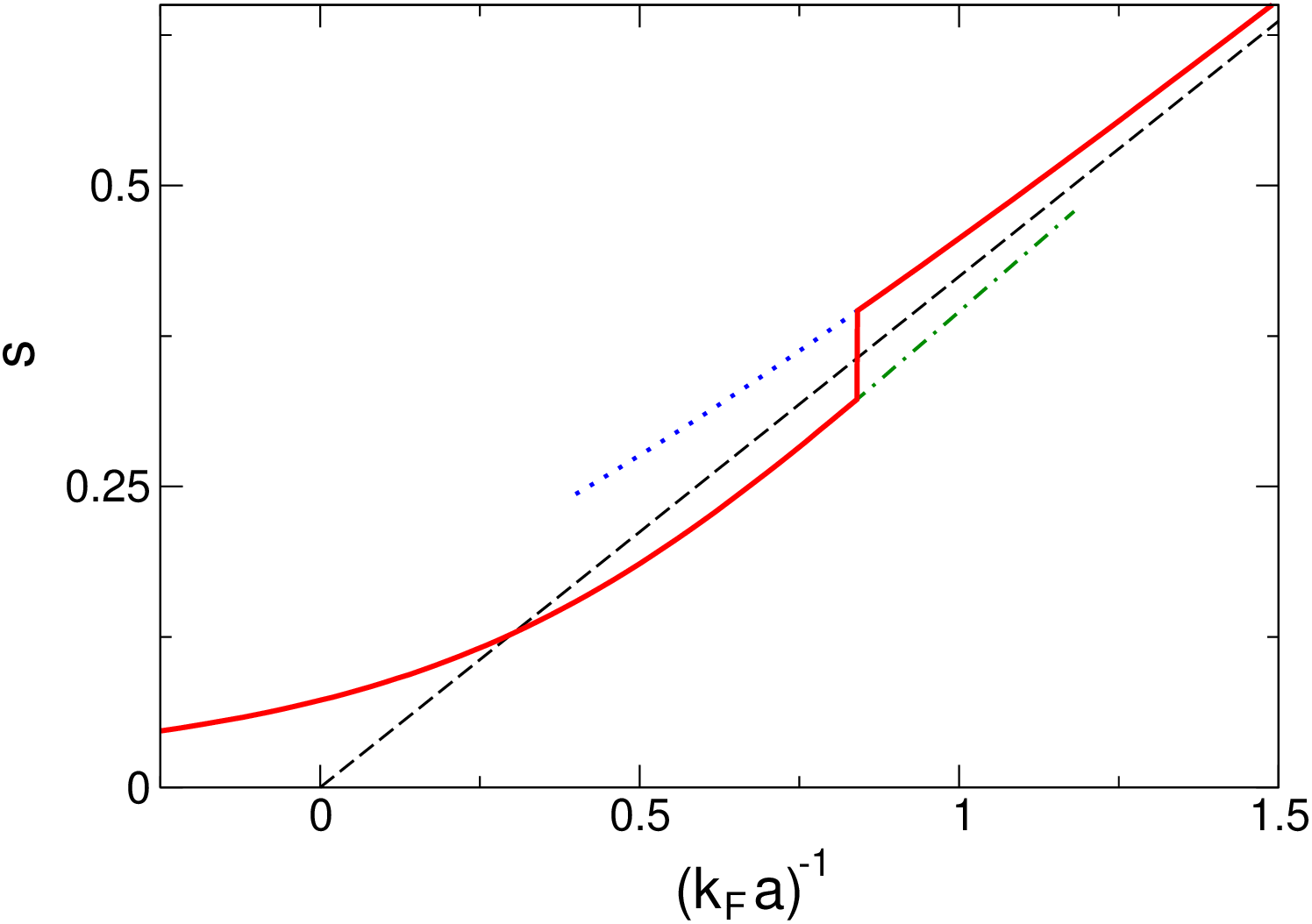}
\caption{(Color online) Dimensionless contact coefficient $s$ as function of
 $(k_F a)^{-1}$, calculated from the two variational wavefunctions \eqref{chevy} and
 \eqref{Varansatz2}. Within this approach $s$ is discontinuous at the critical
 coupling $v_M$. The black dashed line marks the asymptotics in the 
 molecular limit, where $s=4v/3\pi$ is fixed by the two-particle bound state wavefunction
 in momentum space.}
\label{fig:contact}
\end{figure}

A different perspective on the first order nature of the polaron to 
molecule transition is provided by considering the so called contact 
coefficient $C$. As shown by Tan \cite{Tan1}, the momentum distribution of
Fermi gases with zero range interactions generically decays with a power 
law $n_{\sigma}(k)\to C/k^4$ for large momenta. The associated coefficient 
$C$ is identical for both spin components $\sigma=\uparrow, \downarrow$ 
\cite{Randeria09} and is a measure of the probability that two Fermions with opposite spin 
are close to each other \cite{Braaten08}. Using the adiabatic theorem 
derived by Tan \cite{Tan2}, the contact density can be determined
from the derivative 
\begin{equation}
 \frac{\partial \, u}{\partial (1/a)} = - \frac{\hbar^2}{4 \pi m} \, C 
\end{equation}
of the ground state energy density $u=E/V$ with respect to the inverse 
scattering length. Now the definition of the down-spin chemical 
potential $\mu_{\downarrow}$ implies that the energy density $u$  of the strongly 
imbalanced Fermi gas $n_\downarrow \ll n_\uparrow$ to linear order in the minority 
density $n_\downarrow$ is of the form
\begin{equation}
u=\frac{3}{5} \varepsilon_{F\uparrow} n_\uparrow+\mu_\downarrow
 n_\downarrow+ \dots
\end{equation}
where the first term is simply the energy of a non-interacting gas of spin-up Fermions.
The dimensionless contact coefficient $s$ defined by $C=s\cdot k_{F}k_{F\downarrow}^3$ 
for a strongly imbalanced Fermi gas can thus be obtained from the derivative
\begin{equation}
s = \frac{1}{3 \pi} \frac{\partial (-\mu_\downarrow/\varepsilon_{F})}{\partial v}
\end{equation}
of the negative down-spin chemical potential in units of the Fermi energy 
with respect to the coupling constant $v$. Since $\mu_{\downarrow}$ is 
precisely the energy $E$ associated with adding a single down-spin, our 
result for the ground state energy of the $(N+1)$-particle problem
immediately gives the contact density of an almost fully polarized 
attractive Fermi gas (note that this applies even on the molecular side 
$v>v_M$, where the single added down-spin is not a propagating 
quasiparticle). The associated dimensionless constant $s$ 
is shown in Fig. \ref{fig:contact}. It increases monotonically from weak 
coupling to unitarity and up to the critical coupling $v_M$. At this point, 
there is a discontinuous jump upwards, that reflects the transition to a 
molecular state. Note, that the proportionality 
$C\sim k_{F\downarrow}^3\sim n_{\downarrow}$
of the contact to the down-spin density makes $C$ vanish 
in the limit of full polarization. This is expected, because the fully 
polarized gas at zero temperature is an ideal Fermi gas, 
with no tails in the momentum distribution. 
Apart from the jump at $v_M$, the behavior of the dimensionless contact 
coefficient $s$ is rather close to that obtained for the 
contact coefficient $C=s\cdot\tilde{k_F}^4$ of the 
balanced superfluid along the BCS-BEC crossover \cite{Haussmann09}
(note that the Fermi momentum $\tilde{k_F}$ of the balanced gas is
related to that of the up-spin component used here by
$k_F^3=\tilde{k_F}^3(1+\sigma)$, where $\sigma=(n_{\uparrow}-n_{\downarrow})/
(n_{\uparrow}+n_{\downarrow})$ is the degree of polarization at a fixed
total number of particles). 
Indeed, in weak coupling one obtains $s_{wc}=(2/3\pi v)^2$ from the mean
field attraction of the polaron to the up-spin Fermi sea, while 
$s\simeq 0.08$ at unitarity and $s_{BEC}=4v/3\pi$ in the molecular 
limit, very similar to the  behavior that is found for $s$ in the balanced 
superfluid \cite{Haussmann09}.

\begin{figure}
\includegraphics[width=0.95 \columnwidth]{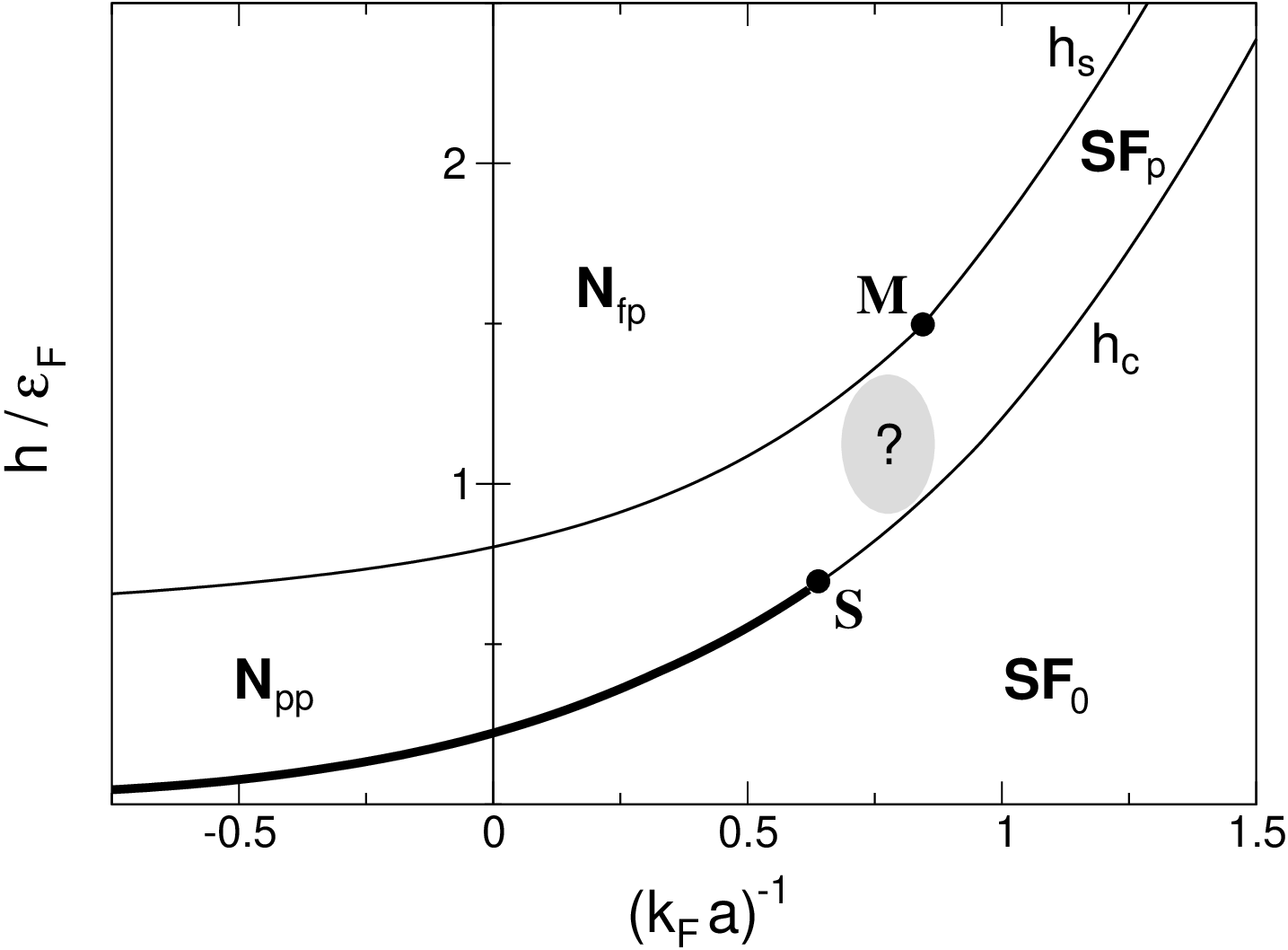}
\caption{Qualitative phase diagram of the imbalanced Fermi gas as a function of the inverse coupling strength
$(k_F a)^{-1}$ and the effective magnetic field $h/\varepsilon_F$. The thick line indicates a first order phase transition and
the different phases are labeled as in \cite{Pilati08}, i.e. $N_{\text{fp}}$: fully polarized normal phase, $N_{\text{pp}}$: partially polarized normal phase, $SF_0$: balanced superfluid, $SF_\text{p}$: polarized superfluid. The points $M$ and $S$ are discussed in the text. The precise structure of the phase diagram in the nontrivial regime $h_c<h<h_s$ is likely to contain unconventional superfluid phases in addition to the $N_{\text{pp}}$ and $SF_\text{p}$ phase, which are not shown in our figure.}
\label{fig:phasdiag}
\end{figure}

The solution of the $(N+1)$-body problem for arbitrary coupling strengths $v$
has also implications for the phase diagram of the imbalanced 
Fermi gas in the regime near complete polarization. For a discussion of this
issue, it is convenient to introduce an effective magnetic
field $h$ that couples to the two-different spin-states $\sigma=\uparrow,\downarrow$
in the standard form 
 \begin{equation}
\hat{H}'=-h\left(\hat{N}_{\uparrow}-\hat{N}_{\downarrow}\right)
\end{equation}
of a 'Zeeman' field that favors a finite population imbalance $\sigma=(n_{\uparrow}-n_{\downarrow})/
(n_{\uparrow}+n_{\downarrow})> 0$. At a fixed total density $n$, the ground state
energy $u$ per volume is then a function of $n$ and $h$. It determines
the chemical potentials of the majority and minority species from $\mu_{\uparrow,\downarrow}
=\mu\pm h$ where $\mu=\partial u(n,h)/\partial n$ is the average chemical potential.
In addition, it also fixes the imbalance from $n_{\uparrow}-n_{\downarrow}=-\partial u(n,h)/\partial h$.  
The choice of an ensemble with fixed values of $n$ and $h$ is convenient
for a discussion of the ground state phase diagram of the attractive Fermi gas 
at arbitrary coupling $v$, both in the homogeneous case and in the presence of
a harmonic trap \cite{Footnote1}. Indeed, there are two critical fields $h_c(v)$ and $h_s(v)$ that
separate two simple limiting phases from a regime, in which nontrivial ground states
are expected: The lower critical field $h_c$ is defined by $\sigma(h)\equiv 0$ for
$h<h_c$ and determines the boundary of the balanced superfluid phase (denoted by $SF_0$
in Fig. \ref{fig:phasdiag}, following the notation used by Pilati and Giorgini \cite{Pilati08}).
The upper critical (or 'saturation') field $h_s$, in turn, is defined by the condition
of complete polarization  $\sigma(h)\equiv 1$ for $h>h_s$. Since a single component
Fermi system has vanishing interactions in the ultracold limit, this regime is
just an ideal Fermi gas, i.e. it is a normal, fully polarized state.  The qualitative 
structure of the zero temperature phase diagram as a function of the interaction parameter
$v=1/(k_F a)$ and the effective magnetic field $h$ in units of the bare Fermi energy
$\varepsilon_F$ of the fully polarized gas is shown in Fig. \ref{fig:phasdiag}.

In this diagram, the upper line $h_s(v)$ is completely fixed by our calculation above of 
the energy $\mu_{\downarrow}$ associated with adding a single down-spin
to an up-spin Fermi sea. Indeed, since $\mu_{\uparrow}\equiv\varepsilon_F$ along 
this line, we have $h_s=(\varepsilon_F-\mu_{\downarrow})/2$. In terms of the
constant $A(v)$ introduced in Eq.  \eqref{E(p)}, this leads to $h_s/\varepsilon_F
=(1-A)/2$,  giving $h_s=0.81\varepsilon_F$ at unitarity from the precise
numerical value of the polaron energy \cite{Prokofev2}. On the molecular side,
Eq. \eqref{asympen} gives 
\begin{equation}
\frac{h_s}{\varepsilon_F} = v^2 + 1 - \frac{a_{ad}/a}{2\pi v} \, + \ldots \ ,
\label{hs}
\end{equation}
which is very accurate even at $v=v_M$. The point $M$ along this line separates 
a regime where a single down-spin is 
a well defined fermionic quasiparticle from one, in which it is bound 
to the up-spin Fermi sea. The first order nature of the transition shows up
as a discontinuity of the slope in $h_s(v)$ at $M$ which is, however,
hardly visible in Fig. \ref{fig:phasdiag}. For a finite density of down-spins, 
the point $M$ appears as an endpoint of a line that separates a phase 
with a finite Fermi surface volume $\Omega_{\downarrow}\ne 0$ to its left
from one with $\Omega_{\downarrow}=0$ \cite{Footnote2}. Using the generalized 
Luttinger theorem derived by Sachdev and Yang \cite{Sachdev06},
the expected polarized superfluid ($SF_p$) phase on the molecular 
side has a condensate of 'dimers' plus an up-spin Fermi sea, whose
volume $\Omega_{\uparrow}=(2\pi)^3(n_{\uparrow}-n_{\downarrow})$ 
is set by the imbalance. This is consistent with the naive picture
that the density of unpaired up-spins is simply $n_{\uparrow}-n_{\downarrow}$
even though the 'dimers' in the vicinity of the transition are far from
local $(\uparrow,\downarrow)$-pairs.  In principle, this simple picture of the $SF_p$- phase
as a BEC coexisting with a sharp, single Fermi surface of unpaired up-spins
is unstable with respect to p-wave pairing due to the induced 
interactions between the unpaired fermions through the superfluid \cite{Bulgac06}.
In practice, the nontrivial superfluid phase of the unpaired up-spins
is exponentially suppressed for strong imbalance. Moreover, quantitative results 
for the p-wave instability can be derived only in second order in $1/v\ll 1$, where
the resulting energy scales are exponentially small compared with $\varepsilon_F$.  
In practice, therefore, the phase with p-wave pairing among the unpaired 
up-spins seems hardly accessible experimentally. 

A nontrivial issue that has been neglected in the discussion so far is the question
whether a gas of polarons or bound molecules is indeed stable at low but finite
densities $n_{\downarrow}$. On the weak coupling side, there is again an 
induced attractive interaction in the p-wave channel among both the up-spins
and the down-spins, mediated by the other species. The ground state is
thus expected to be a two-component p-wave superfluid and not a normal Fermi
liquid state \cite{Bulgac06}. Similar to the situation in the BEC-limit, however
the energy scale for this instability is exponentially small in the regime where
the calculation can be controlled. More importantly, as has been shown 
recently by Nishida \cite{Nishida09}, the effective interaction between two heavy 
down-spin fermions immersed in an up-spin Fermi sea is attractive in the 
p-wave channel only for weak coupling. Approaching unitarity, the p-wave 
interaction becomes repulsive. Assuming that this result carries over to the 
relevant case of equal masses of the up- and down-spin Fermions, 
a finite density gas of down-spins will indeed form a normal Fermi liquid
at unitarity, as was implicitly assumed in the calculations of the equation of
state and density profiles of the unitary gas beyond the critical imbalance 
$\sigma_c\simeq 0.4$, where the balanced superfluid is no longer stable
\cite{Lobo06, Giorgini08}. On the molecular side, the phase immediately
below the saturation field line $h_s(v)$ is expected to be a superfluid of
$(\uparrow,\downarrow)$-pairs at a very low density $n_{\downarrow}\to 0$
immersed in an up-spin Fermi sea. The fact that the atom-dimer repulsion
$a_{ad}=1.18\, a$ is much larger than the dimer-dimer repulsion $a_{dd}=0.6\, a$
\cite{Petrov05}, however indicates that a low density gas of molecules
tends to phase separate from the up-spin Fermi gas. This phase separation
has indeed been found from an extended BCS-description of the BCS-BEC 
crossover in an imbalanced gas \cite{bedaque, Sheehy, Yip06, Parish07}. It has recently
been seen also in the variational Monte Carlo calculations by Pilati and 
Giorgini \cite{Pilati08}. Their results indicate that a section between $v_N\simeq 0.73$ and a 
triple point at $v_T\simeq 1.7$ along the $h_s$-line is actually a first order 
line, where the polarized superfluid disappears with a finite jump in density
as the effective field $h$ increases through $h_s$.
As shown above, the point $M$ lies in the interval between $v_N$ and $v_T$ and thus the
polaron to molecule transition would not be accessible at any finite minority density,
at least not in an equilibrium situation. Clearly, our variational calculation for the single down-spin
problem cannot address the question of phase separation.
An unexpected feature of the $h_s$-line
in the presence of phase separation is the fact that the transition across $h_s$
is predicted to be continuous up to $v_N$, first order between $v_N$ and $v_T$ and
continuous again for $v>v_T$.  The rather large value $v_T\simeq 1.7$
up to which phase separation is predicted also appears surprising.
Indeed, in this regime a mean field theory describing a 
Fermi gas coexisting with a BEC of molecules gives for the energy 
per volume as a functional of the density difference $\delta n=n_{\uparrow}-
n_{\downarrow}$ and the dimensionless field  $\tilde{h}=h/\varepsilon_F$ the simple form
\begin{eqnarray}
\frac{u(n,\, \delta n,\, h)}{\varepsilon_F} &=& \frac{3}{5} \Big( \frac{\delta n}{n}\Big)^{2/3} \delta n \, - \, v^2 (n-\delta n)  \notag \\
&+& \frac{\tilde{a}_{ad}}{2 \pi v} \frac{(n-\delta n)\, \delta n}{n} + \frac{\tilde{a}_{dd}}{12 \pi v} \frac{(n-\delta n)^2}{n} \notag \\
 &-& \tilde{h} \, \delta n 
\label{LGS}
\end{eqnarray}
Here, $\tilde{a}_{ad}=a_{ad}/a$ and $\tilde{a}_{dd}=a_{dd}/a$ are the atom-dimer and dimer-dimer scattering length measured in units of the atom-atom scattering length and $v=(k_F a)^{-1}$.  
The true ground state energy density $u(n,h)$ as function of the total density $n$ and the effective magnetic field $h$ is determined by the minimum of the Landau energy \eqref{LGS} 
with respect to the density imbalance 
\begin{eqnarray}
\frac{\partial (u/\varepsilon_F)}{\partial \, \delta n} &=& \Big( \frac{\delta n}{n}\Big)^{2/3} + \, v^2 \, + \frac{\tilde{a}_{ad}}{2 \pi v} \Big( 1- 2 \frac{\delta n}{n}\Big) \notag \\
&+& \frac{\tilde{a}_{dd}}{6 \pi v} \Big(\frac{\delta n}{n}-1\Big) - \tilde{h} \, \overset{!}{=}\,  0
\label{GS}
\end{eqnarray}
This equation determines the imbalance $\sigma=\delta n /n$ as a function of the 
field $h$ and indeed, it correctly describes the exact asymptotic results for both the saturation field
$h_s$ and the lower critical field $h_c$ in the limit $v\gg 1$ (see  Eqs. \eqref{hs} and \eqref{hc}). In this simple model,
phase separation between a polarized superfluid phase and a fully polarized Fermi gas appears for coupling constants  
$v<v_{c,PS}$, below which the energy density \eqref{GS} has a second minimum  
at full polarization $\sigma\equiv 1$. This occurs at 
\begin{equation}
v_{c,PS}=\frac{3}{2 \pi} \Big( \tilde{a}_{ad}-\frac{\tilde{a}_{dd}}{6} \Big)\, .
\end{equation}
With the exact values $\tilde{a}_{ad}=1.18$ and $\tilde{a}_{dd}=0.6$ one obtains 
$v_{c,PS}=0.516$, where the simple expansion \eqref{LGS}, however, is no longer valid.
From the calculation above, the triple point $v_T$ beyond which phase separation appears in an almost fully polarized
gas lies at a much smaller value of the coupling strengths than found previously \cite{Parish07,Pilati08}.
At finite temperatures, phase separation is suppressed by the presence of a mixing entropy, which may explain that it is not observed in the experiments, where $T \approx 0.15 T_F$.

Concerning the lower critical field $h_c(v)$, its weak-coupling limit
is determined by the well known Chandrashekar-Clogston result $h_c = \Delta/\sqrt{2}$,
beyond which the balanced BCS-pairstate is unstable \cite{Chandrasekhar, Clogston}.
In the molecular limit $v\gg 1$, the critical field 
\begin{equation}
\frac{h_c}{\varepsilon_F}=v^2+\frac{1}{2\pi va}\left(a_{ad}-\frac{a_{dd}}{6}\right)+\ldots
\label{hc}
\end{equation}
follows essentially the two-particle
binding energy with corrections due to the atom-dimer and dimer-dimer scattering
lengths $a_{ad}$ and $a_{dd}$, respectively.  At unitarity, $h_c=0.26\varepsilon_F$ is
a universal constant times the bare up-spin Fermi energy $\epsilon_F$ \cite{Lobo06}.
As a result, there is a wide range $h_s/h_c = 3.12$ between the balanced superfluid and the fully polarized gas, much larger than that found in an $N=\infty$ theory of the imbalanced attractive Fermi gas, where $h_s/h_c = 1.24$ \cite{Nikolic}.
For $h>h_c$ the balanced superfluid is destroyed by the onset of 
a finite polarization $\sigma\ne 0$, which leads to a mismatch of the Fermi energies.
An effective field theory due to Son and Stephanov \cite{Son06} indicates, that the phase
beyond the balanced superfluid exhibits a spatially varying superfluid order of the FFLO type
as also found at weak coupling. The transition is first order with a jump both in polarization
$\sigma$ an total density, a situation that is also found in the case of a direct transition between
a balanced superfluid and a partially polarized normal phase \cite{Lobo06, Giorgini08}.
In an ensemble with a given density that is used here, the $h_c$-line would then split into two
distinct lines $h_{c1}$ and $h_{c2}$, as noted by Sheehy and Radzihovsky \cite{Sheehy}. In our
diagram in Fig.\ \ref{fig:phasdiag}, $h_c$ denotes the boundary of the balanced superfluid at given
density, which is well defined without specifying the state that is reached at nonzero polarization.
The first order nature of the transition is found only up to a splitting point
$S$, beyond which the fermionic excitations have their minimum at $\mathbf{p}=0$. 
When this is the case, additional up-spin Fermions can be added by filling up
a Fermi surface whose volume $\Omega_{\uparrow}\sim\sigma\sim
(h-h_c)^{3/2}$ increases continuously from zero. The transition
 from the balanced superfluid to a polarized superfluid with unpaired excess 
Fermions is therefore continuous and preserves superfluidity. The precise location
of the splitting point S has been determined recently from a calculation of the
fermionic excitation spectrum along the BCS-BEC crossover of the balanced
gas \cite{Haussmann09}. It is located at $v_S\simeq 0.63$ and $h_c(v_S)=\Delta \simeq 0.6 \varepsilon_F$, at considerably larger
coupling strengths than predicted by mean-field theory where the splitting
point coincides with the zero crossing of the chemical potential (note the factor
$2^{1/3}$-difference with the result in Ref. \cite{Haussmann09}, which is due
to the fact that the up-spin Fermi wave vector and not that of 
the balanced case appears in our present coupling constant $v$). 
This is in agreement with the calculation of the splitting point within an $\epsilon-4-d$-expansion
by Nishida and Son \cite{Nishida07} but is larger than the value $v_S\simeq 0.5$
found for the splitting point in the Monte Carlo calculations of
Pilati and Giorgini \cite{Pilati08}. Note that the possibility of extracting
the critical coupling $v_S$ of the splitting point from a calculation of the
balanced gas relies on the fact that the ground state energy is independent
of the field $h$ in the whole regime $h\leq h_c$ because the polarization
$\sigma=-n\partial u(n,h)/\partial h$ vanishes. The nature of the phase diagram
near the splitting point has been discussed by
Son and Stephanov \cite{Son06} using an effective field theory. In particular, 
the phase immediately beyond $h_c$ is expected to be of the FFLO-type,
with a spatially oscillating superfluid order parameter that appears also
in weak coupling beyond the Chandrasekhar-Clogston limit \cite{Footnote3}.
It is an open question, how this nontrivial superfluid evolves into 
a normal  phase in which the two spin components each form a Fermi liquid.
In fact it is this latter phase, which describes the experimentally 
observed density profiles \cite{Shin08} at unitary extremely well \cite{Lobo06, Giorgini08}.
It is also an open issue, of how to separate in detail the regime between the lower 
critical field and the saturation field into a regime where an imbalanced Fermi liquid
or a polarized superfluid phase appear as ground states. In the phase diagram of
Pliati and Giorgini, the first order line that bounds the balanced superfluid 
up to the splitting point S extends as a first order line up to $h_s$ at the 
coupling $v_N$ and then continues along $h_s$ up to the tricritical point $v_T$.

\section{Conclusions}
From a variational wavefunction that describes the $(N+1)$-particle problem
of a single down-spin interacting strongly with an up-spin Fermi sea, 
we have discussed the physics of the strongly imbalanced Fermi gas. 
In particular, we have focused our attention on the quasiparticle residue
and the contact coefficient $C$. The latter exhibits a discontinuous jump 
at the polaron to molecule transition which might be detected by measurements of
the closed channel fraction similar to the analysis of the experiments
by Partridge et.al. \cite{Partridge05} due to Werner, Tarruell and Castin \cite{Werner09}.
A motivation for this work were the recent experiments of Schirotzek et.al. \cite{Schirotzek09},
who have observed a transition from a Fermi liquid phase of polaronic 
quasiparticles near unitarity to a phase in which the quasiparticle residue
vanishes. As shown above, this transition is expected to be a 
discontinuous one in the single down-spin limit. Apparently, however,
$Z_{\downarrow}$ vanishes in a continuous manner in the experiment
(see Fig. \ref{fig:Zchevy}).  Apart from the uncertainty
in extracting $Z_{\downarrow}$ from the sharp structure in the minority 
rf-spectrum, this discrepancy 
is probably due to the fact that the Chevy wavefunction \eqref{chevy}
strongly overestimates the quasiparticle residue $Z_\downarrow$ near the polaron to molecule transition.
A reliable quantitative calculation of the rf-spectra for both 
finite concentrations and at finite temperatures is, unfortunately, not available.
The existence of a stable finite density gas of polarons in the regime up
to $v\simeq 0.9$ however indicates that the interaction between them is repulsive,
so that they indeed form a Landau Fermi liquid below the critical coupling $v_M$.
As discussed in section IV, the detailed structure of how this phase connects
to the nontrivial superfluid phases expected near the splitting point S and on the
BEC-side is a major and still open problem.

\acknowledgments
\noindent

We acknowledge many useful discussions with T. Enss, A. Recati and M. Zwierlein. 
Moreover, we would like to thank A. Schirotzek, N. Prokof'ev and B. Svistunov for providing 
their experimental respectively their numerical results for
comparison. W. Z. is grateful to W. Ketterle and M. Zwierlein for their hospitality 
at the MIT-Harvard Center for Ultracold Atoms during a sabbatical,
where this work was started.

\emph{Note added:} The variational wavefunction \eqref{Varansatz2} for the $(N+1)$-particle
problem in the molecular regime has been found independently by
C.\ Mora and F.\ Chevy  \cite{ChevyMora}. For the calculation of the ground state energy
they have assumed a vanishing hole wavevector $\mathbf{q}=0$, which reduces
the resulting integral equation to a one-dimensional problem.

Furthermore, we are grateful to R.\ Combescot for pointing out their closely related work \cite{CGL}.

\appendix*

\section{three particle limit}
\label{appendix_A}

In the following we briefly show, how the exact solution of the three Fermion problem can be obtained from the integral equation
\eqref{intequ} for the full variational wavefunction \eqref{ansatz2}. If the (N-1)-particle Fermi sea is reduced to a single $\uparrow$-Fermion, only the $\q=0$ terms remain in the variational wavefunction \eqref{ansatz2}. Thus, starting from the integral equation \eqref{intequ} in the thermodynamic limit, taking the limit $k_F \rightarrow 0$ and setting $\q = 0$, one arrives at the simplified equation
\begin{equation}
\alpha_{\k 0} \, \eta_{\k 0} = -\int \frac{d^3k'}{(2 \pi)^3} \frac{\eta_{\k' 0}}{E_{\k' \k 0}} \ .
\label{eqA1}
\end{equation}
Inserting the coefficients $\alpha_{\k 0}$ and $E_{\k' \k 0}$ from Eqs.\ \eqref{coeff_E3} and \eqref{coeff_alpha} explicitly, the integral equation \eqref{eqA1} takes the form
\begin{eqnarray}
\bigg( \frac{1}{a} &-& \sqrt{\frac{3 k^2}{4} - m E} \, \bigg) \eta_\k \, = \notag \\
&&= \int \frac{d^3k'}{(2 \pi)^3} \frac{4 \pi \, \eta_{\k'}}{k^2 + k'^2 + \k \cdot \k'-mE} \ . \label{eqA2}
\end{eqnarray}
Note that $\eta_\k \equiv \eta_{\k 0}$ corresponds to the Fourier transform of the relative wavefunction between the ($\uparrow,\downarrow$)-molecule and the additional $\uparrow$-Fermion.
The integral equation \eqref{eqA2} is exactly the same as the one obtained by Skorniakov and Ter-Martirosian \cite{Skorniakov57} for the three Fermion problem. In particular it is equivalent to Equ.\ (29) in \cite{Skorniakov57}, which corresponds to three-nucleon scattering with total isospin $T=1/2$ and total spin $S=3/2$ (note that the spin part only contributes an unimportant prefactor to the wavefunction in this case).


\begin{thebibliography}{999}
\bibitem{Footnote1}Within a local density approximation, the different phases that
appear in a trap with a spatially dependent coupling constant $v(\mathbf{x})$ due
to the decrease of the local Fermi wavevector $k_F(\mathbf{x})$ from the center of
the trap to its edge simply follow from a parabolic (vertical at unitarity) line in Fig. \ref{fig:phasdiag}
 \begin{equation}
\frac{h}{\varepsilon_F(\mathbf{x})}=\frac{2h}{|E_b|}\cdot v^2(\mathbf{x})\, ,
\end{equation}
where $|E_b|=1/(ma^2)$ for both positive or negative scattering lengths $a$. 
Its curvature $2h/|E_b|$ is fixed by the global imbalance,
while the initial point $v(\mathbf{x}=0)$ is determined by the local 
Fermi wavevector at the trap center $\mathbf{x}=0$. Note that in a trap 
different phases are spatially separated for both continuous and first order
transition lines. By contrast, genuine phase separation is associated
with a first order transition and appears only in the homogeneous system.
\bibitem{Footnote2}Note the down-spin Fermi surface volume $\Omega_{\downarrow}$
vanishes also along the $h_s$-line to the left of $M$. This continuous transition, however, 
becomes a smooth crossover at any finite temperature. By contrast, the transition
between a normal and a superfluid state that are both partially polarized, also 
remains at finite temperature.
\end{thebibliography}
\end{document}